\documentclass[%
superscriptaddress,
reprint,
 amsmath,amssymb,
 aps, physrev,
prb,
]{revtex4-2}

\usepackage{graphicx}
\usepackage{dcolumn}
\usepackage{bm}

\usepackage{xcolor}
\usepackage{color}

\begin{document}


\title{\textbf{Dimensional crossover of class D real-space topological invariants} 
}%

\author{Martin Rodriguez-Vega}
\email{rodriguezvega.physics@gmail.com}
\affiliation{American Physical Society, 1 Physics Ellipse Dr, College Park, MD, USA}

\author{Terry A. Loring}
\email{tloring@unm.edu}
\affiliation{Department of Mathematics and Statistics, University of New Mexico, Albuquerque, NM,
USA}

\author{Alexander Cerjan}
\email{awcerja@sandia.gov}
\affiliation{Center for Integrated Nanotechnologies, Sandia National Laboratories, Albuquerque, NM,
USA}

\date{\today}
\begin{abstract}
The topological properties of a material depend on its symmetries, parameters, and spatial dimension. Changes in these properties due to parameter and symmetry variations can be understood by computing the corresponding topological invariant. Since topological invariants are typically defined for a fixed spatial dimension, there is no existing framework to understand the effects of changing spatial dimensions via invariants. Here, we introduce a framework to study topological phase transitions as a system's dimensionality is altered using real-space topological markers. Specifically, we consider Shiba lattices, which are class D materials formed by magnetic atoms on the surface of a conventional superconductor, and characterize the evolution of their topology when an initial circular island is deformed into a chain. We also provide a measure of the corresponding protection against disorder. Our framework is generalizable to any symmetry class and spatial dimension, potentially guiding the design of materials by identifying, for example, the minimum thickness of a slab required to maintain three-dimensional topological properties.
\end{abstract}

\maketitle

\section{\label{sec:Intro} Introduction}

The possible topological properties of non-interacting Hamiltonians are dictated by their spatial dimension, and discrete and crystal symmetries \cite{RevModPhys.88.035005}. In particular, changing the dimension of a system while keeping its discrete symmetries unaltered generally yields a topological phase transition, following the Altland-Zirnbauer classes \cite{PhysRevB.55.1142,schnyder2008,10.1063/1.3149495,Ryu_2010}. Examples in condensed matter systems include transitions from a trivial insulator to a quantum spin Hall (QSH) phase in HgTe quantum wells \cite{doi:10.1126/science.1133734, doi:10.1126/science.1148047}, and transitions from three-dimensional topological insulators with two-dimensional surface states to QSH states with one-dimensional helical states \cite{Zhang2010, Park2015,  doi:10.1021/acs.nanolett.9b01641, doi:10.1021/acs.nanolett.6b02044}. 

Such topological transitions due to dimensional shifts have been previously theoretically characterized by analyzing edge states in finite-size systems, the topological invariants based on band representations \cite{PhysRevB.81.041307, Scheurer2015}, and the transport properties of a material \cite{PhysRevB.92.235407}. However, a real-space framework capable of illuminating how a system's topology changes during a dimensional shift has traditionally been missing due to the dearth of local topological markers needed to probe the topology of systems lacking translational symmetry across all of the possible symmetry classes. Nevertheless, such a theory could be useful to guide the design of materials and next-generation devices; for example, to identify how thin a slab can be while maintaining its three-dimensional topological properties, or how thick a nanowire can be while maintaining its end Majorana modes.

\begin{figure}[th]
    \centering
    \includegraphics[width=1\linewidth]{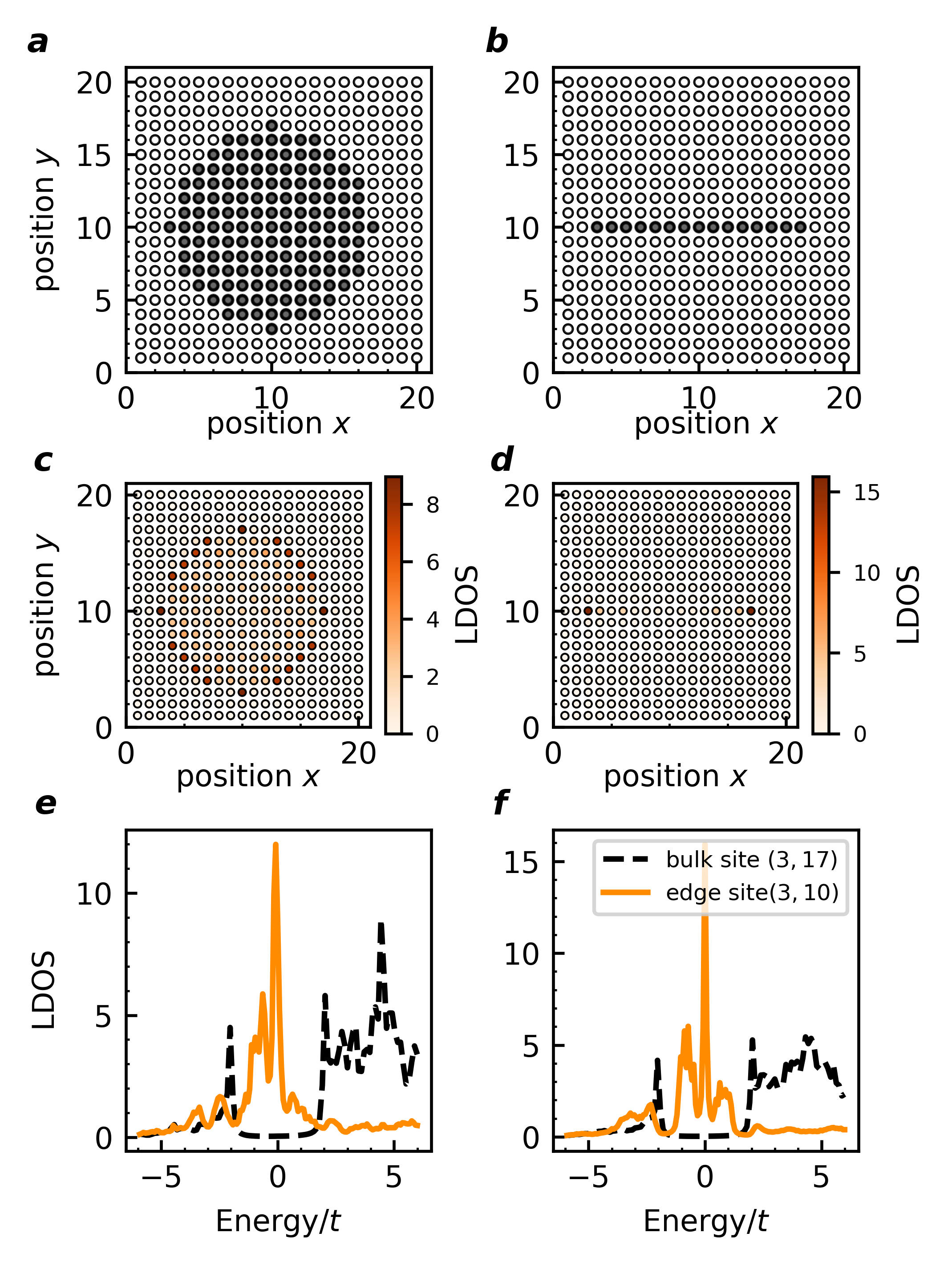}
    \caption{\textbf{Topological phenomena in 1D and 2D Shiba lattices.} \textbf{a},\textbf{b} Magnetic adatoms (filled circles) arranged in a circular island (\textbf{a}) and a chain (\textbf{b}) geometry, embedded in a two-dimensional superconductor whose sites are represented by empty circles. \textbf{c},\textbf{d} Zero-energy local density of states (LDOS) for the circular Shiba island and chain, respectively, showing a Majorana mode at their boundary. \textbf{e},\textbf{f}  LDOS as a function of energy at a bulk site (dashed black line) and at a site on the island's and chain's edges (solid orange line). The latter displays zero-energy peaks. The parameters used for the island and chain are $N=20$, where $N$ is the size of the superconductor's lattice, chemical potential $\mu=-4t$, spin-orbit coupling strength $\alpha=0.45t/l$, superconducting s-wave pairing $\Delta=2.0t$, and magnetic moment times exchange coupling $JS=4.8t$.}
    \label{fig:fig1}
\end{figure}

In this paper, we introduce a scheme to study topological phase transitions induced by the change in a system's spatial dimensions via local topological markers. We will focus on the symmetry class D in the Altland-Zirnbauer classification \cite{PhysRevB.55.1142,schnyder2008,10.1063/1.3149495,Ryu_2010}, a class that contains some topological superconductors \cite{Kitaev_2001}, and which exhibits non-trivial invariants in both 1D and 2D. Specifically, Hamiltonians in this class have particle-hole symmetry $\mathcal{C}$, and broken time-reversal and chiral symmetries. In one dimension (1D), the topological invariant corresponds to $\mathbb{Z}_2$, while in two dimensions (2D) to $\mathbb{Z}$. Our aim is to study the evolution of the $\mathbb{Z}$ invariant into $\mathbb{Z}_2$ as one of the dimensions of the target material embedded in an insulating environment is reduced.

In particular, we consider magnet-superconductor hybrid systems, Shiba lattices, consisting of time-reversal-symmetry-breaking magnetic atoms placed on the surface of a superconductor arranged as 1D chains \cite{PhysRevB.84.195442, PhysRevB.88.020407} or in 2D lattices \cite{PhysRevLett.114.236803, Li2016}. Shiba lattices can be assembled with high precision in the laboratory and have shown signatures of topological superconductivity. 1D Shiba chains have been fabricated using ferromagnetic transition metal atoms (such as iron, chromium, manganese, and cobalt) placed on the surface of lead, rhenium, niobium or $\beta$-Bi$_2$Pd \cite{Nadj_Perge_2014, PhysRevLett.115.197204, Pawlak2016, Feldman2017, doi:10.1126/science.aan3670, Ruby2017, doi:10.1126/sciadv.aar5251,Schneider2020,PhysRevB.104.045406,Schneider2021,Schneider2022}. In 2D, Shiba lattices have been realized using iron adatoms on rhenium \cite{doi:10.1126/sciadv.aav6600}, flakes of chromium tribromide placed on niobium diselenide \cite{Kezilebieke2020,Kezilebieke2022}, and cobalt-silicon islands on Si(111) covered by lead \cite{Menard2017}. Additionally, arrays of chromium atoms on niobium have shown signatures of mirror symmetry-protected topological superconductivity \cite{Soldini2023}. 

The task of tracking the topological properties of the Shiba lattice across spatial dimensions is well suited to the spectral localizer framework \cite{LORING2015383, loring2017finite, Loring_2020}, a real-space numeric $K$-theory approach that characterizes the local topological properties of finite-size systems, without requiring any notion of lattice periodicity. The spectral localizer approach has been employed to identify topological phenomena in 
gapless systems \cite{Schulz-Baldes_2021,PhysRevB.106.064109, PhysRevB.109.195107,Cheng2023}, nonlinear materials \cite{PhysRevB.108.195142}, non-Hermitian systems \cite{Cerjan_2023,liu_mixed_2023,ochkan_non-hermitian_2024,chadha_real-space_2024,garcia2024clifford}, photonic systems \cite{CerjanLoring+2022+4765+4780, PhysRevLett.131.213801, Wong2024}, quasicrystals \cite{PhysRevLett.116.257002,saito2024gapless,10.1063/1.5083051}, time-dependent systems \cite{ghosh2024local,PhysRevB.110.014309},
disordered superconductors \cite{zakharov2024majoranametal} and heterostructures \cite{PhysRevLett.132.073803}. 
While most of these studies have looked at 1D and 2D systems, the framework has also been applied to a 3D system \cite{LORING2015383} and a 4D system \cite{cerjan2024even}.

For the present study, we utilize two local topological markers derived from the spectral localizer, a Chern number, and a $\mathbb{Z}_2$ invariant that characterizes two- and one-dimensional Shiba lattices embedded in two-dimensional superconductors. Additionally, we also considered the one-to-zero-dimensional crossover in a BDI class model,  showing the generality of our approach (See Supplementary Note~1). As this framework does not rely on projecting to an occupied subspace, it is endowed with a measure of the corresponding local topological protection related to the local spectral gap \cite{cerjan2025detectinglocaltopologyspectral}. The local nature of the markers permits us to track their changes and corresponding local protection as the dimension of the Shiba lattice changes from two to one dimension. The real-space topology in 2D Shiba lattices has been characterized using a global position-space invariant \cite{PhysRevB.100.235102,PhysRevB.100.184510}, and $\mathbb{Z}_2$ local markers for nonchiral phases in odd dimensions have been recently introduced \cite{PhysRevLett.129.277601}. However, these approaches cannot diagnose the topology of 1D Shiba lattices embedded in a two-dimensional superconductor.


\section{Results}
\subsection{\label{sec:shiba} Shiba lattice model Hamiltonian}

A mean-field tight-binding Hamiltonian describing Shiba lattices, systems consisting of a set of magnetic atoms placed on the surface of a superconductor, is given by \cite{Li2016, PhysRevB.100.184510}
\begin{align}
    H & = -t \sum_{\mathbf{r}, \boldsymbol{\delta}, \sigma}  c^\dagger_{\mathbf{r}\sigma} c_{\mathbf{r}+\boldsymbol{\delta}\sigma}  - \mu \sum_{\mathbf{r},\sigma}  c^\dagger_{\mathbf{r}\sigma} c_{\mathbf{r}\sigma} + \Delta \sum_{\mathbf{r}} c^\dagger_{\mathbf{r} \uparrow} c^\dagger_{\mathbf{r} \downarrow} + h.c. \nonumber \\ 
    & + i \alpha \sum_{\mathbf{r}, \boldsymbol{\delta},a,b} c^\dagger_{\mathbf{r} a} \left(\boldsymbol{\delta} \times \boldsymbol{\sigma} \right)^z_{ab} c_{\mathbf{r}+\boldsymbol{\delta}b} + J \sum_{\mathbf{R},a,b} \mathbf{S} \cdot c^\dagger_{\mathbf{R}a} \boldsymbol{\sigma}_{ab} c_{\mathbf{R}b},
    \label{eq:H}
\end{align}
where $c^\dagger_{\mathbf{r}\sigma}$ ($c_{\mathbf{r}\sigma}$) creates (annihilates) an electron at site $\mathbf{r}$ with spin $\sigma$, $\boldsymbol{\sigma}=\{\sigma_x, \sigma_y, \sigma_z\}$ are the Pauli matrices, $t$ is the hopping amplitude between nearest-neighbor lattice sites separated by the displacement vector $\boldsymbol{\delta}$ such that the distance between nearest-neighbor sites is $l=|\boldsymbol{\delta}|$, $\mu$ is the chemical potential, $\alpha$ is the spin-orbit coupling strength, and $\Delta$ is the superconducting s-wave pairing. Specifically, we consider a square lattice superconductor. The magnetic adatoms located at positions $\mathbf{R}$ are assumed single-orbital, and are described as classical magnetic moments $\mathbf{S}$ aligned perpendicular to the superconductor lattice plane and coupled to the electron spin via the exchange coupling $J$.  

Shiba lattices possess particle-hole symmetry $\mathcal{C}^2 = I$, where $I$ is the identity, such that they are in class D of the Altland-Zirnbauer classification and can thus exhibit non-trivial topology in physically realizable dimensions. To illustrate the topological behavior of Shiba lattices, we consider two magnetic adatom geometries in Fig.~\ref{fig:fig1}: a circular Shiba island (a)  and a chain (b), both embedded in a two-dimensional square-lattice superconductor. We compute the local density of states (LDOS) at site $i$ and energy $E$,
\begin{equation}
\rho_i(E) = -\frac{1}{\pi} \sum_{\sigma} \textrm{Im} \left[ G^{\sigma \sigma}_{i i}(E) \right]
\label{eq:ldos}
\end{equation}
using the Chebyshev--Bogoliubov--de Gennes method \cite{PhysRevLett.105.167006}, where $G^{\sigma \sigma}_{ii}(E)$ is the Green's function, and $\sigma$ corresponds to the spin. We include an artificial broadening of the Green’s function’s features at zero temperature [see Methods]. Majorana modes localized at the corresponding 1D and 0D edges of the circular island and chain are numerically observed in their respective zero-energy LDOS,  see Fig.~\ref{fig:fig1}(c) and (d). Moreover, we compare the LDOS as a function of energy at bulk and edge sites of the magnetic adatom geometries in Fig.~\ref{fig:fig1}(e) and (f); the former case shows the superconducting gap, while the latter exhibits a zero-energy peak corresponding to the interface-localized states. 

Previously, topological properties of finite-sized two-dimensional Shiba lattices have been classified in real-space using Chern markers \cite{PhysRevB.100.235102}, including hybrid structures composed of networks of chains attached to islands \cite{PhysRevB.100.184510}. Topological quantum numbers for one-dimensional chains based on the scattering matrix have been introduced~\cite{PhysRevB.84.195442}. Similarly, scattering theory has been recently applied to diagnose intrinsic higher order topology protected by spatial symmetries in finite samples \cite{zijderveld2025scatteringtheoryhigherorder}. A real-space topological invariant for one-dimensional time-quasiperiodic Majorana modes has been discussed in Ref.~\cite{PhysRevB.110.014309}.

\subsection{\label{sec:SpectralLocalizer} Spectral localizer}

Here, we use the spectral localizer framework to study the experimentally relevant geometry of a one-dimensional Shiba chains embedded in a two-dimensional superconductor, such as the case in Fig.~\ref{fig:fig1}(d). In general, this framework has two key features that make it amenable to this task: first, it possesses local topological markers for both 1D and 2D class D systems, and second, it is endowed with a local measure of the associated topological protection.
Physically, the spectral localizer framework assesses whether a system can be locally continued to an atomic limit while preserving any local spectral gap and (relevant) symmetries, or if there is an obstruction to doing so.
Mathematically, the spectral localizer framework classifies material topology by first translating a system's description into a single Hermitian matrix and then diagnoses its topology via matrix homotopy, i.e., the study of which types of invertible matrices can be path-connected such that every matrix element in the path shares the same properties (e.g., Hermiticity) and remains invertible. As such, the spectral localizer framework does not rely on vector bundles, though it is provably equivalent in some cases \cite{loring2017finite,jezequel2025}.
For a detailed tutorial on the spectral localizer and Clifford pseudospectrum, see Ref.~\cite{Cerjan_2024}.

For a $d$-dimensional system with open boundaries described by the Hamiltonian $H$ defined on a finite-dimensional Hilbert space, the spectral localizer is defined as \cite{LORING2015383}
\begin{align}
    L_{\boldsymbol{\lambda}  } &(\boldsymbol{X},H)  =  \sum_{j=1}^d \kappa (X_j - x_j I) \otimes \Gamma_j+ 
    \left(H - E I\right) \otimes \Gamma_{d+1}, \label{eq:Ldef}
\end{align}
where the matrices $\Gamma_i$ generate an irreducible Clifford representation, $X_j$ are the position operators, $\boldsymbol{\lambda} = (\mathbf{x},E)$ is the point in position-energy space at which we are probing the local topological properties, and $\kappa>0$ is a scaling parameter with units of energy divided by length. Intuitively, the spectral localizer can be understood as combining the eigenvalue equations of the position operators with the Hamiltonian into a single matrix using a Clifford representation to preserve the independent information contained in these different operators. Thus, $\kappa$ must be  chosen to balance the relative contributions of $X_j$ and $H$ to the spectrum of $L_{\boldsymbol{\lambda}}$; choosing $\kappa$ to be too large or too small results in the information carried by $ L_{\boldsymbol{\lambda}}$ to effectively duplicate the information already present in $X_j$ or $H$, respectively. For bounded Hamiltonians, such as those describing tight-binding models, a proven range of validity for $\kappa$ exists where the spectral localizer provides information beyond what is immediately available in $H$ and $\boldsymbol{X}$ \cite{Loring_2020,cerjan2025detectinglocaltopologyspectral,jezequel2025}. Generally, for systems with gapped bulk spectra, this scaling parameter can be set by the bulk gap width $E_{\textrm{gap}}$ and the smallest length of the finite system $L$ as $\kappa \sim E_{\textrm{gap}}/L$. In comparison with other theories of local topological markers \cite{kitaev2006anyons,bianco_mapping_2011}, setting $\kappa$ is similar to the choice of integration radius used by those markers.


The spectral localizer $L_{\boldsymbol{\lambda}} (\boldsymbol{X},H)$ encodes both the local topological invariant and a measure of its local protection. In particular, as this framework's local markers are all rooted in matrix homotopy arguments, the spectral localizer quantifies the strength of the local topological protection by how close a system's $L_{\boldsymbol{\lambda}  } (\boldsymbol{X},H)$ is to being non-invertible at a chosen $\boldsymbol{\lambda}$. More precisely, the localizer gap is defined as
\begin{equation}
 \mu_{\boldsymbol{\lambda}}(\boldsymbol{X},H) =   \text{min}\left\{ \left|\Sigma\left( L_{\boldsymbol{\lambda}} (\boldsymbol{X},H)\right) \right| \right\},
\end{equation}
where $\Sigma\left( L_{\boldsymbol{\lambda}} (\boldsymbol{X},H)\right)$ corresponds to the spectrum of the spectral localizer. In other words, this localizer gap is the minimum distance an eigenvalue must move for $L_{\boldsymbol{\lambda}  }$ to become non-invertible. As the spectral localizer is Hermitian (for Hermitian $H$), any perturbation $\delta L$ to $L_{\boldsymbol{\lambda}}$, either through changing the choice of $\boldsymbol{\lambda}$ or perturbing the underlying system $H \rightarrow H + \delta H$, cannot yield a non-invertible $L_{\boldsymbol{\lambda}} + \delta L$ so long as $\Vert \delta L \Vert < \mu_{\boldsymbol{\lambda}}$. Here, $\Vert \cdot \Vert$ is the $L_2$ matrix norm, i.e., its largest singular value.

The localizer gap can also be viewed as a quantity similar to the local density of states. As the spectral localizer is a combination of generally non-commuting operators, $\boldsymbol{\lambda} = (\mathbf{x},E)$ can be understood as a guess at a location in position and energy space where the operators might exhibit an approximate joint eigenvector $|\phi\rangle$. If $\mu_{\boldsymbol{\lambda}}$ is small relative to quantities like $\sqrt{\Vert[H,\kappa X_j]\Vert}$, the system is guaranteed to exhibit a state that satisfies $X_j|\phi\rangle \approx x_j |\phi\rangle$ and $H|\phi\rangle \approx E |\phi\rangle$ \cite{cerjan_V_L2023quadraticPS}. Thus, altogether, large values of $\mu_{\boldsymbol{\lambda}}$ indicate locations where the local topology is robust, while small values of $\mu_{\boldsymbol{\lambda}}$ indicate the presence of approximate states in the underlying system. 

\begin{figure}
    \centering
    \includegraphics[width=1\linewidth]{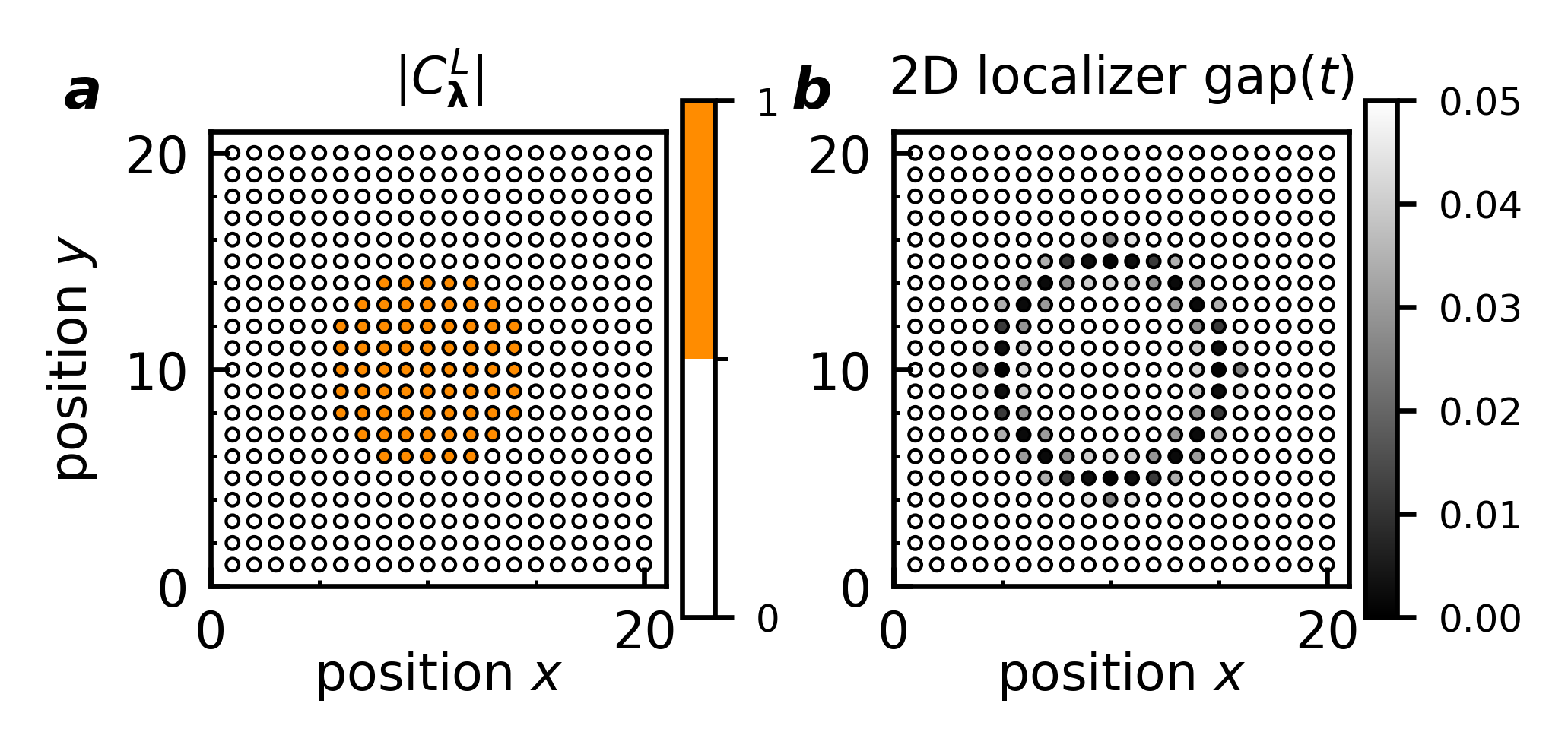}
    \caption{\textbf{Application of the spectral localizer framework to the Shiba island.} \textbf{a},\textbf{b} Chern number $|C_{\boldsymbol{\lambda}}^\textrm{L}|$ (\textbf{a}) and 2D localizer gap (\textbf{b}) as a function of position for the 2D island geometry shown in Fig.~\ref{fig:fig1}\textbf{a}, for the same parameters and at $E/t=0$. The spectral localizer calculations use $\kappa = 0.05(t/l)$.
    }
    \label{fig:fig2dtop}
\end{figure}

\subsection{2D class D local invariant}

While the definitions of the spectral localizer and the associated localizer gap can be defined for arbitrary dimensions, the associated local markers for classifying material topology require specializing to both the specific system dimensionality and symmetry class being considered. This demand for specificity can both be viewed as a physical consequence of the Altland-Zirnbauer classification \cite{schnyder2008,10.1063/1.3149495,Ryu_2010}, as well as a mathematical consequence of changing dimensions and symmetries altering the structure of $L_{\boldsymbol{\lambda}} (\boldsymbol{X},H)$. For example, such alterations may transform a spectral localizer that is symmetric $L_{\boldsymbol{\lambda}}^\top = L_{\boldsymbol{\lambda}}$ into one that no longer possesses this property, fundamentally changing what matrix homotopy arguments can be applied and thus changing the structure, and possible existence, of the local markers.

For 2D systems, the Pauli matrices can be chosen as the Clifford representation in Eq.~\eqref{eq:Ldef}, yielding
\begin{align}    L^{(2D)}_{\boldsymbol{\lambda}  =(x,y,E)} &(X,Y,H)  =  \left(H - E I\right) \otimes \sigma_z \nonumber \\ & + \kappa \left[ (X - x I) \otimes \sigma_x + (Y - y I) \otimes \sigma_y  \right].
\label{eq:L2d}
\end{align}
For a Hamiltonian in class D, the local topological marker at the energy-position point $\boldsymbol{\lambda}  =(x,y,E)$ is \cite{LORING2015383} 
\begin{equation}
    C_{\boldsymbol{\lambda}}^\textrm{L} = \frac{1}{2} \text{sig}\{ L^{(2D)}_{\boldsymbol{\lambda}}\left(X,Y,H \right)\}, \label{eq:cherndef}
\end{equation}
where $\text{sig}\{ L^{(2D)}_{\boldsymbol{\lambda}}\left(X,Y,H \right)\}$ corresponds to the signature of the spectral localizer, the difference between the number of positive and negative eigenvalues. Thus, when $\mu_{\boldsymbol{\lambda}}$ is nonzero, $C_{\boldsymbol{\lambda}}^\textrm{L}$ takes a well-defined integer value. Note that $C_{\boldsymbol{\lambda}}^\textrm{L}$ can only change its value across $\boldsymbol{\lambda}$ where $\mu_{\boldsymbol{\lambda}}=0$, as the eigenvalues of Hermitian operators must move continuously.

The overall picture of the spectral localizer framework is shown in Fig.~\ref{fig:fig2dtop}, where the local marker $|C_{\boldsymbol{\lambda}}^\textrm{L}|$ and localizer gap $\mu_{\boldsymbol{\lambda}}$ are both shown over the spatial extent of the full 2D system. In the lattice's center the local marker reveals non-trivial topology and the localizer gap is large, indicating the robustness of the topological phase. As the choice of position is varied towards the system's edge, the localizer gap closes around the boundary of the distribution of magnetic adatoms, beyond which the local index changes its value to reveal trivial topology. Finally, at the heterostructure's boundary, the closing of the localizer gap coincides with the chiral interface-localized states, see Fig.~\ref{fig:fig1}c. Thus, bulk-boundary correspondence manifests in the spectral localizer framework as the local index cannot change without crossing through a location in position-energy space where $\mu_{\boldsymbol{\lambda}}=0$, guaranteeing the appearance of a system state at that approximate energy and position.

\begin{figure*}
    \centering
    \includegraphics[width=0.9\linewidth]{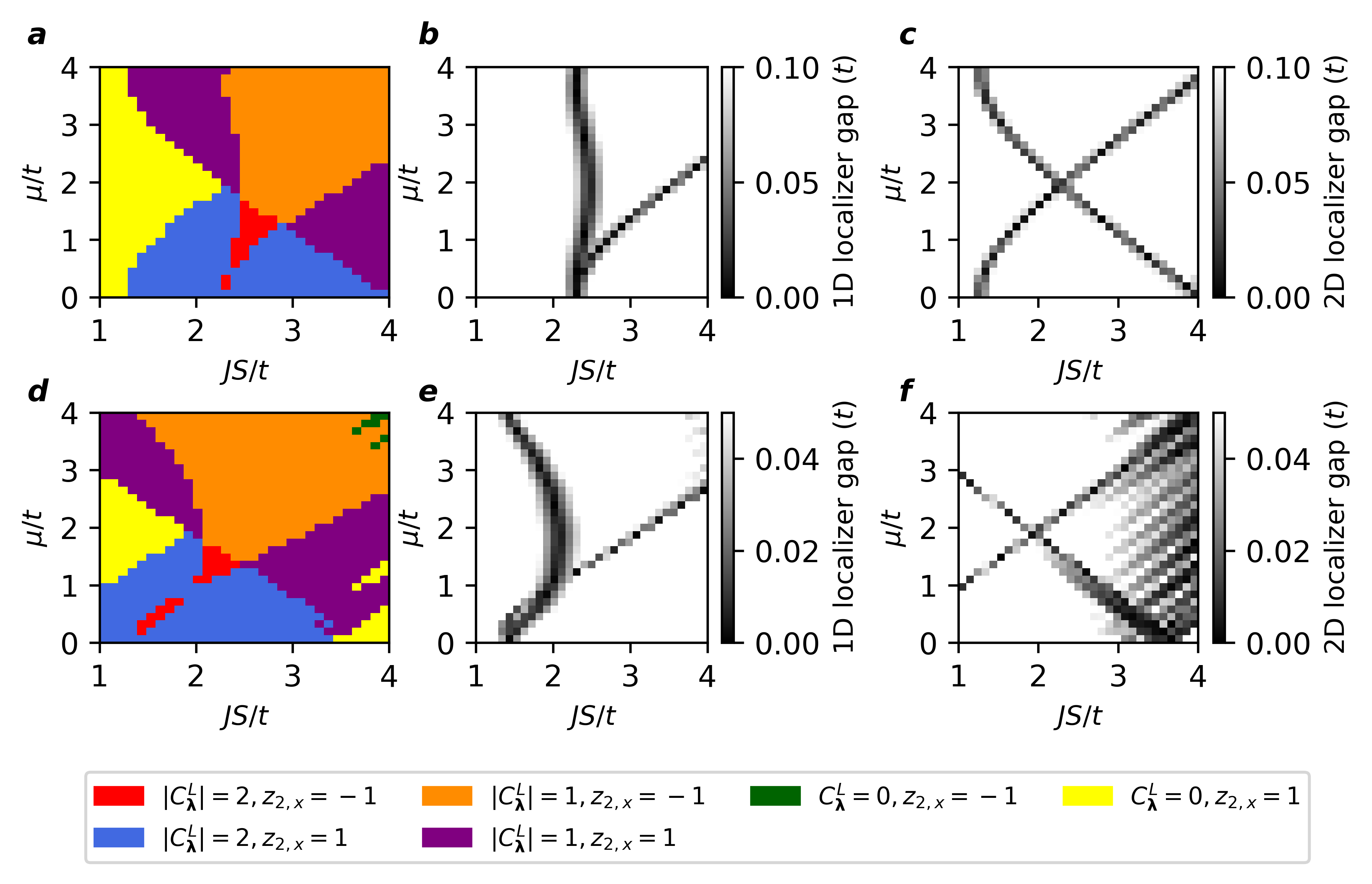}
    \caption{\textbf{Dimensional crossover topological phase diagram in the Shiba lattice.} The phase diagram is shown as a function of the chemical potential $\mu/t$ and the product of the magnetic moment and exchange coupling $JS/t$, for $N=20$, and spin-orbit coupling strength $\alpha=0.45t/l$. The phase diagrams contain $31 \times 31 $ points. \textbf{a}-\textbf{c} Phase diagram (\textbf{a}), 1D spectral localizer gap for the chain (\textbf{b}), and 2D spectral localizer gap for the chain (\textbf{c}) for $\Delta=1.2 t$ at $E=0$. The color code indicates the 2D local marker $|C_{\boldsymbol{\lambda}}^\textrm{L}|$ computed at the center of a circular island, and the 1D local marker $z_{2,x}$ computed at the center of a chain using the same parameters and only changing the geometry. The localizer gaps are similarly computed at the center of the 2D island and 1D chain. \textbf{d}-\textbf{f} Similar to \textbf{a}-\textbf{c}, except for $\Delta=0.4t$. Topological transitions not captured in the diagram can occur as the geometry changes from 2D to 1D. The spectral localizer calculations use $\kappa = 0.05 (t/l)$. 
    }
    \label{fig:phase-diagram}
\end{figure*}

\begin{figure*}
    \centering
    \includegraphics[width=0.9\textwidth]{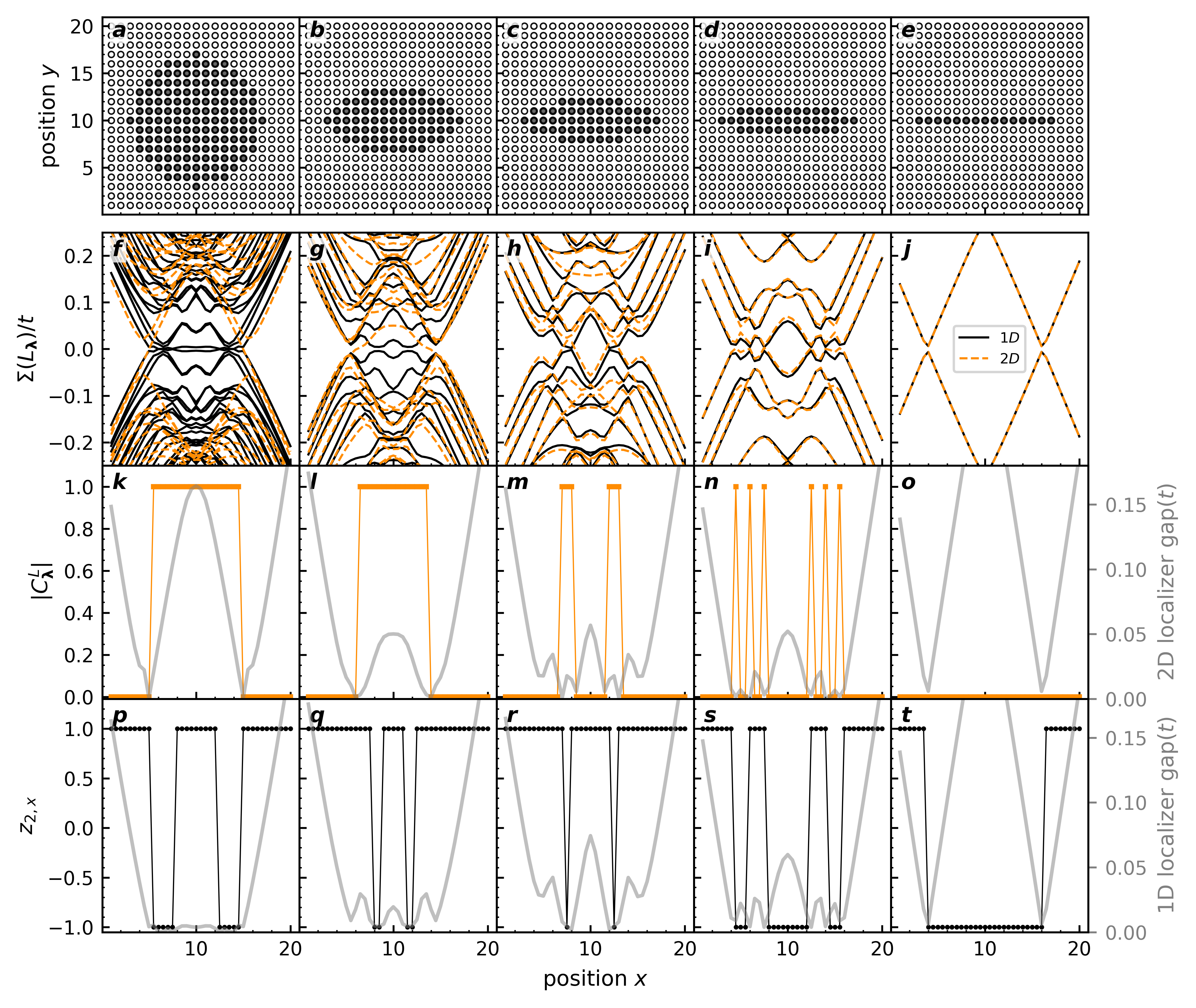}
    \caption{\textbf{Topological phase transitions due to dimensional crossover between the Shiba island and the Shiba chain.} The parameters used are the same as in Fig.~\ref{fig:fig1}, $\Delta =2.0t$, $\mu = -4t$, $\alpha = 0.45t/l$, and $JS = 4.8t$, such that both the initial Shiba island and final Shiba chain exhibit non-trivial topology. \textbf{a}-\textbf{e} Arrangements of magnetic adatoms (filled circles) and their absence (empty circles) embedded in a two-dimensional superconductor interpolating from the two-dimensional circular island to the one-dimensional chain geometries. \textbf{f}-\textbf{j} 1D (black solid line) and 2D (dashed orange lines) localizer spectrum $\Sigma(L_{\boldsymbol{\lambda}})$ as a function of position $x$ for $y = N/2$ and $E=0$. \textbf{k}-\textbf{o} Local Chern number $|C_{\boldsymbol{\lambda}}^\textrm{L}|$ (orange squares) and 2D localizer gap $\text{min}\{| \Sigma(L^{(2D)}_{\boldsymbol{\lambda}})| \}$ (solid gray line) as a function of position $x$ for $y = N/2$ and $E=0$. \textbf{p}-\textbf{t} Local $z_{2,x}$ invariant (black circles) and 1D localizer gap $\text{min}\{| \Sigma(L^{(1D)}_{\boldsymbol{\lambda}}) | \}$ (solid gray line) as a function of position $x$ for $E=0$. The spectral localizer calculations use $\kappa = 0.05(t/l)$. 
    }
    \label{fig:fig2}
\end{figure*}

\begin{figure*}
    \centering
    \includegraphics[width=1\linewidth]{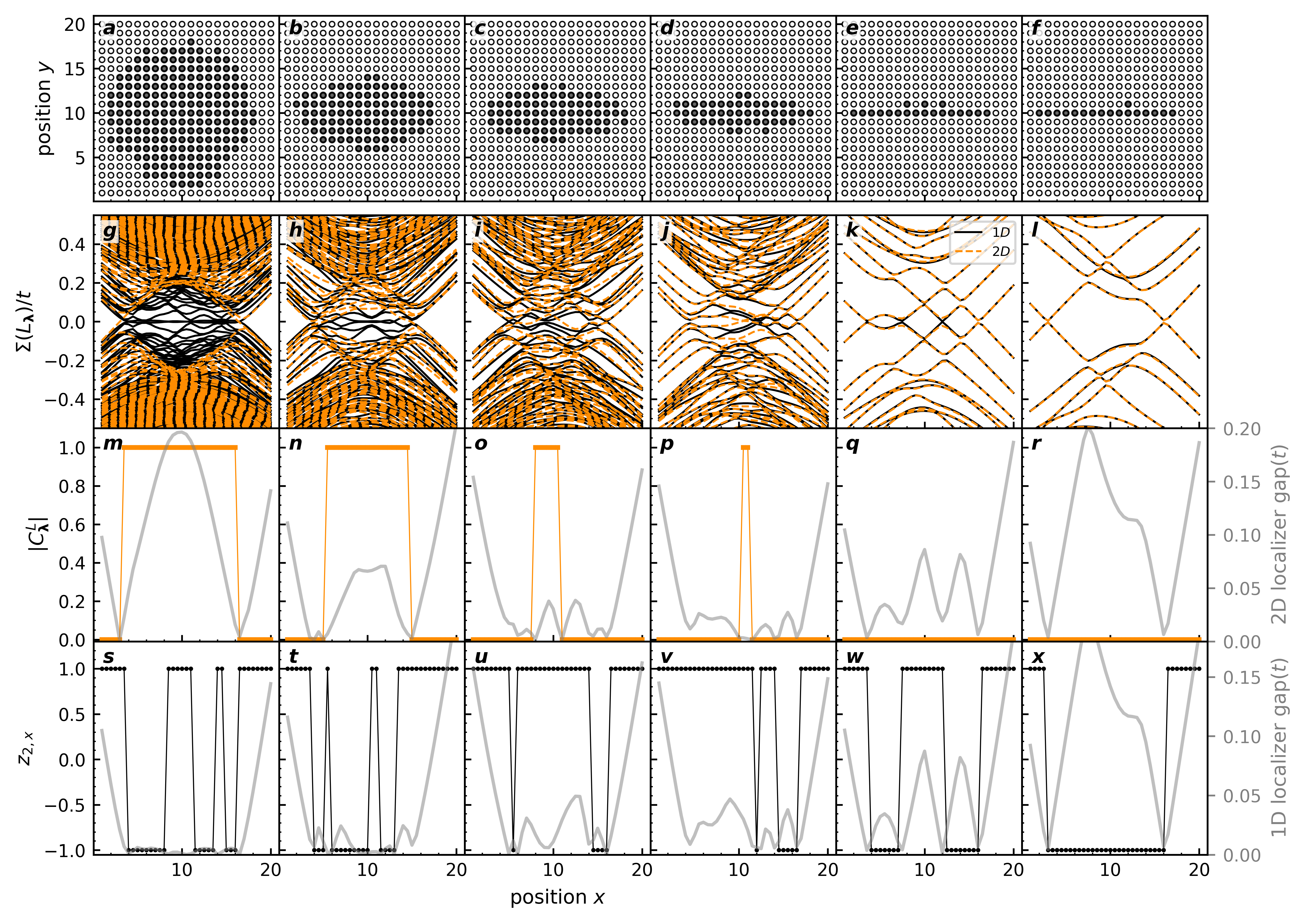}
    \caption{\textbf{Topological phase transitions due to dimensional crossover between an imperfect Shiba island and Shiba chain.} The parameters used are $\mu=-4t$, $\alpha=0.45t/l$, $\Delta=2.0t$, and $JS = 4.8t$, as in Fig.~\ref{fig:fig1}. \textbf{a}-\textbf{f} Arrangements of magnetic adatoms (filled circles) and their absence (empty circles) embedded in a two-dimensional superconductor interpolating from a two-dimensional island with no spatial symmetries to a 1D chain with imperfections. \textbf{g}-\textbf{l} 1D (black solid line) and 2D (dashed orange lines) localizer spectrum $\Sigma(L_{\boldsymbol{\lambda}})$ as a function of position $x$ for $y = N/2$ and $E=0$. \textbf{m}-\textbf{r} Local Chern number $|C_{\boldsymbol{\lambda}}^\textrm{L}|$ (orange squares) and 2D localizer gap $\text{min}\{| \Sigma(L^{(2D)}_{\boldsymbol{\lambda}})| \}$ (solid gray line) as a function of position $x$ for $y = N/2$ and $E=0$. \textbf{s}-\textbf{x} Local $z_{2,x}$ invariant (black circles) and 1D localizer gap $\text{min}\{| \Sigma(L^{(1D)}_{\boldsymbol{\lambda}}) | \}$ (solid gray line) as a function of position $x$ for $E=0$. The spectral localizer calculations use parameter $\kappa = 0.05(t/l)$. 
    }
    \label{fig:random}
\end{figure*}

\subsection{1D class D local invariant}

The 1D spectral localizer can also be constructed using the Pauli matrices as the choice of Clifford representation as
\begin{equation}
L^{(1D)}_{\boldsymbol{\lambda}=(x,E) } (X,H)  =  \left(H - E I\right) \otimes \sigma_y + \kappa (X - x I) \otimes \sigma_x, \label{eq:L1d}
\end{equation}
where $H$ is also given by Eq.~\eqref{eq:H}. Here, the choice of $H$ need not change between using $L^{(2D)}_{\boldsymbol{\lambda}}$ and $L^{(1D)}_{\boldsymbol{\lambda}}$, that is, the same $H$ could be used in both Eqs.~\eqref{eq:L2d} and \eqref{eq:L1d} to determine different physical properties. Instead, to classify topological phenomena associated with a lower physical dimension we use dimensional reduction, which in the spectral localizer framework is performed through the omission of (in this case) the second position operator $Y$ in the definition of $L^{(1D)}_{\boldsymbol{\lambda}}$. Intuitively, the absence of a position operator means that sites with different $y$ coordinates but the same $x$ coordinate are being treated mathematically as analogous to different orbitals of a single site with the same $x$ coordinate, a process that could be considered as similar to the reverse of using a synthetic dimension \cite{Yuan:18} where different degrees of freedom on a single spatial site are considered to constitute an additional fictitious spatial dimension. As such, $L^{(1D)}_{\boldsymbol{\lambda}}$ can be equally applied to 1D chains as well as effective 1D chains embedded in 2D materials (or any material of arbitrarily higher dimension). Similarly, one can also define $L^{(1D)}_{\boldsymbol{\lambda}}$ in terms of other position operators such as $Y$ or $X+Y$. However, for concreteness, we arrange our Shiba chains along the $x$-direction in the 2D superconductor as shown in Fig.~\ref{fig:fig1}b. 

The 1D local topological invariant in class D is defined by first transforming to a basis for which $H$ is purely imaginary and $X$ is real. Such a basis must exist due to the system's particle-hole symmetry. In this basis for $H$, and noting the choice of Pauli matrices used in Eq.~\eqref{eq:L1d}, the 1D spectral localizer is both block off-diagonal and real symmetric at $E=0$, such that $(L^{(1D)}_{(x,0)})^\dagger = (L^{(1D)}_{(x,0)})^\top = L^{(1D)}_{(x,0)}$. Thus, all of its eigenvalues come in pairs, and when $L^{(1D)}_{(x,0)}$ is invertible, its matrix homotopy can be classified using only one of its off-diagonal blocks (which removes this spectral pairing) as \cite{LORING2015383}
\begin{equation}
    z_{2,x} = \text{sign}\left[\text{det} \left\{ \kappa (X-xI) - iH \right\} \right].
\end{equation}
As the determinant is given by the product of a matrix's eigenvalues, $z_{2,x}$ cannot change its value (for fixed $H$) without first crossing through a location in position-space where $\mu_{(x,0)}=0$. This reinforces the fact that $\mu_{\boldsymbol{\lambda}}$ provides both a measure of robustness as well as guarantee of bulk-boundary correspondence regardless of dimension and symmetry class. Note though, that locations where $\mu_{\boldsymbol{\lambda}}(X,H) = 0$ are not generally connected to locations where $\mu_{\boldsymbol{\lambda}}(X,Y,H) = 0$.

\begin{figure*}
    \centering
    \includegraphics[width=0.9\textwidth]{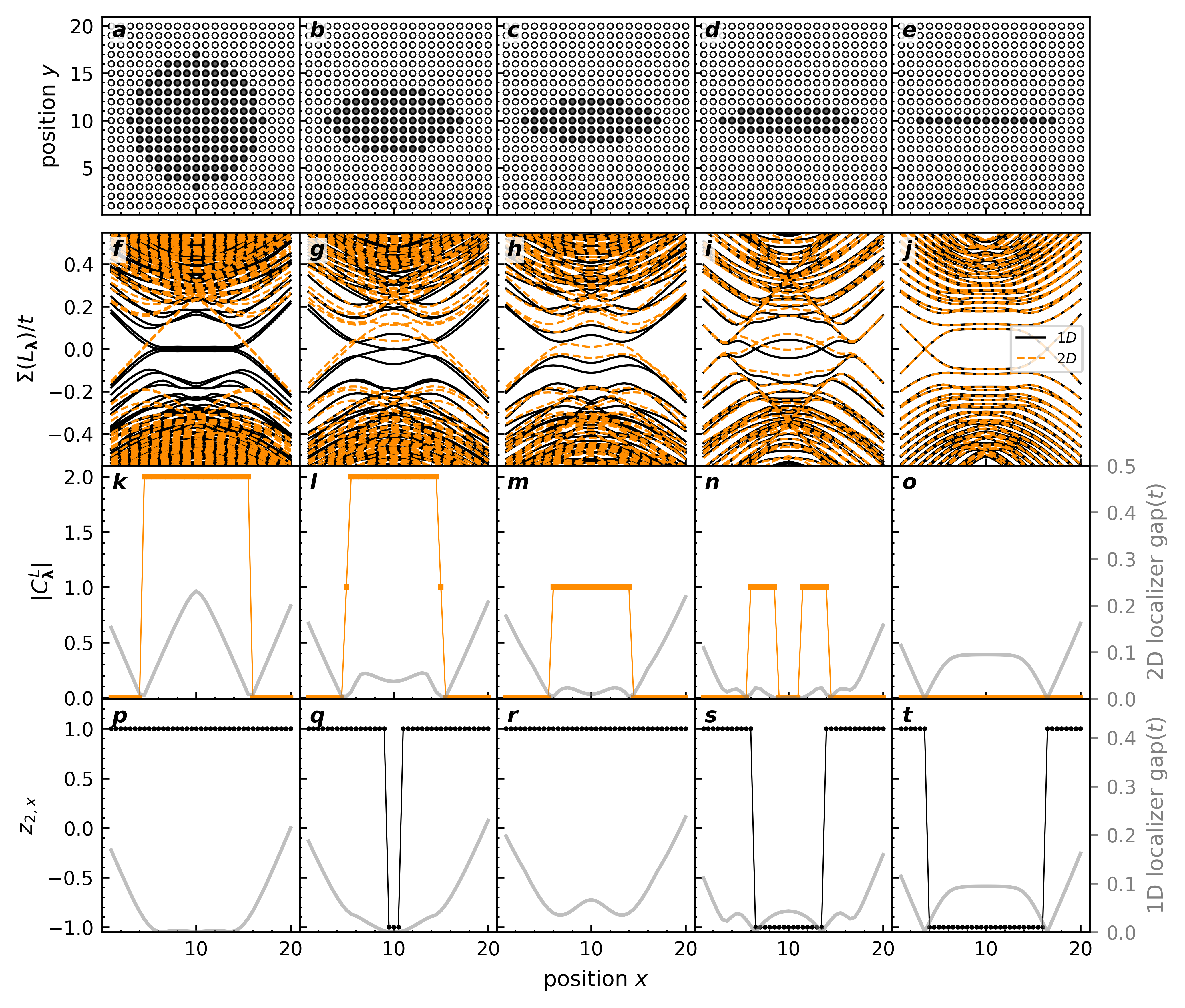}
    \caption{\textbf{Topological phase transitions due to dimensional crossover between a Shiba island with $|C_{\boldsymbol{\lambda}}^\textrm{L}| =2$ and Shiba chain with $z_{2,x}=-1$.} The parameters are $N=20$, $\mu=1.0t$, $\alpha=0.45t/l$, $\Delta=1.2t$, and $JS=2.579t$. \textbf{a}-\textbf{e} Arrangements of magnetic adatoms (filled circles) and their absence (empty circles) embedded in a two-dimensional superconductor interpolating from the two-dimensional circular island to the one-dimensional chain geometries. \textbf{f}-\textbf{j} 1D (black solid line) and 2D (dashed orange lines) localizer spectrum $\Sigma(L_{\boldsymbol{\lambda}})$ as a function of position $x$ for $y = N/2$ and $E=0$. \textbf{k}-\textbf{o} Local Chern number, $|C_{\boldsymbol{\lambda}}^\textrm{L}|$, (orange squares) and 2D localizer gap $\text{min}\{| \Sigma(L^{(2D)}_{\boldsymbol{\lambda}}) | \}$ (solid gray line) as a function of position $x$ for $y = N/2$ and $E=0$. \textbf{p}-\textbf{t} Local $z_{2,x}$ invariant (black circles) and 1D localizer gap $\text{min}\{| \Sigma(L^{(1D)}_{\boldsymbol{\lambda}}) | \}$ (solid gray line) as a function of position $x$ for $E=0$. The spectral localizer calculations use $\kappa = 0.05 (t/l)$. 
    }
    \label{fig:fig3}
\end{figure*}

\subsection{\label{sec:results} Shiba lattice dimensional crossover phase diagram}

Leveraging the local topological invariants for Class D systems, we explore the Shiba lattice's parameter space for 1D and 2D geometries and consider values previously used in the literature for this model \cite{PhysRevB.102.104501, PhysRevB.100.235102, PhysRevB.100.214504, PhysRevB.100.184510}.  To this end, we consider two geometries, a circular island, and a chain, and compute the corresponding local topological invariants (the Chern number $C_{\boldsymbol{\lambda}}^\textrm{L}$ and $z_{2,x}$, respectively) at the system's center, see Fig.~\ref{fig:phase-diagram}. Using the same parameters for both geometries, we find regions where the Shiba island and chain are both topological, regions where the Shiba island is topological and the chain is trivial, regions where the island is trivial and the chain is topological, and regions where both geometries are trivial. Note that topological transitions not captured in the diagram can occur as the geometry changes from 2D to 1D. For example, the transition from $|C_{\boldsymbol{\lambda}}^\textrm{L}|=2$ to $z_{2,x} = -1$ as the circular island is flattened occurs via a change to $|C_{\boldsymbol{\lambda}}^\textrm{L}|=1$ for a narrow island, as we discuss later [see Sec.~\ref{sec:examples} and Fig.~\ref{fig:fig3}]. In Fig.~\ref{fig:phase-diagram}, we consider only positive values for the chemical potential $\mu$, as the phase diagram is symmetric under $\mu \rightarrow -\mu$. In Supplementary Note 2, we present a lower-resolution phase diagram that considers negative values for the chemical potential $\mu$.

Even for the relatively modest system sizes considered in Fig.~\ref{fig:phase-diagram}, the predicted topological phases of the Shiba island and chain are robust, as the localizer gaps [Figs.~\ref{fig:phase-diagram}b,c,e,f] in both dimensions attain values within an order of magnitude of half the system's bulk spectral gap [Figs.~\ref{fig:fig1}e,f]. Note, in an infinite clean system, the localizer gap at mid-gap is expected to be approximately half the bulk spectral gap width, i.e., the distance in energy to the nearest bulk states, as that is the perturbation to $L_{\boldsymbol{\lambda}}$ needed to find an approximate eigenstate. Indeed, there is a broad range of $\kappa$ spanning multiple orders of magnitude that produce quantitatively similar results, see Supplementary Note 3. Thus, the phase diagrams in Fig.~\ref{fig:phase-diagram}a,d are also robust to the choice of position where the topology is determined, so long as this choice is a few sites from the Shiba geometry's boundary.

\subsection{Topological phase transitions from dimensional crossover \label{sec:examples}}

\begin{figure}
    \centering
    \includegraphics[width=0.9\linewidth]{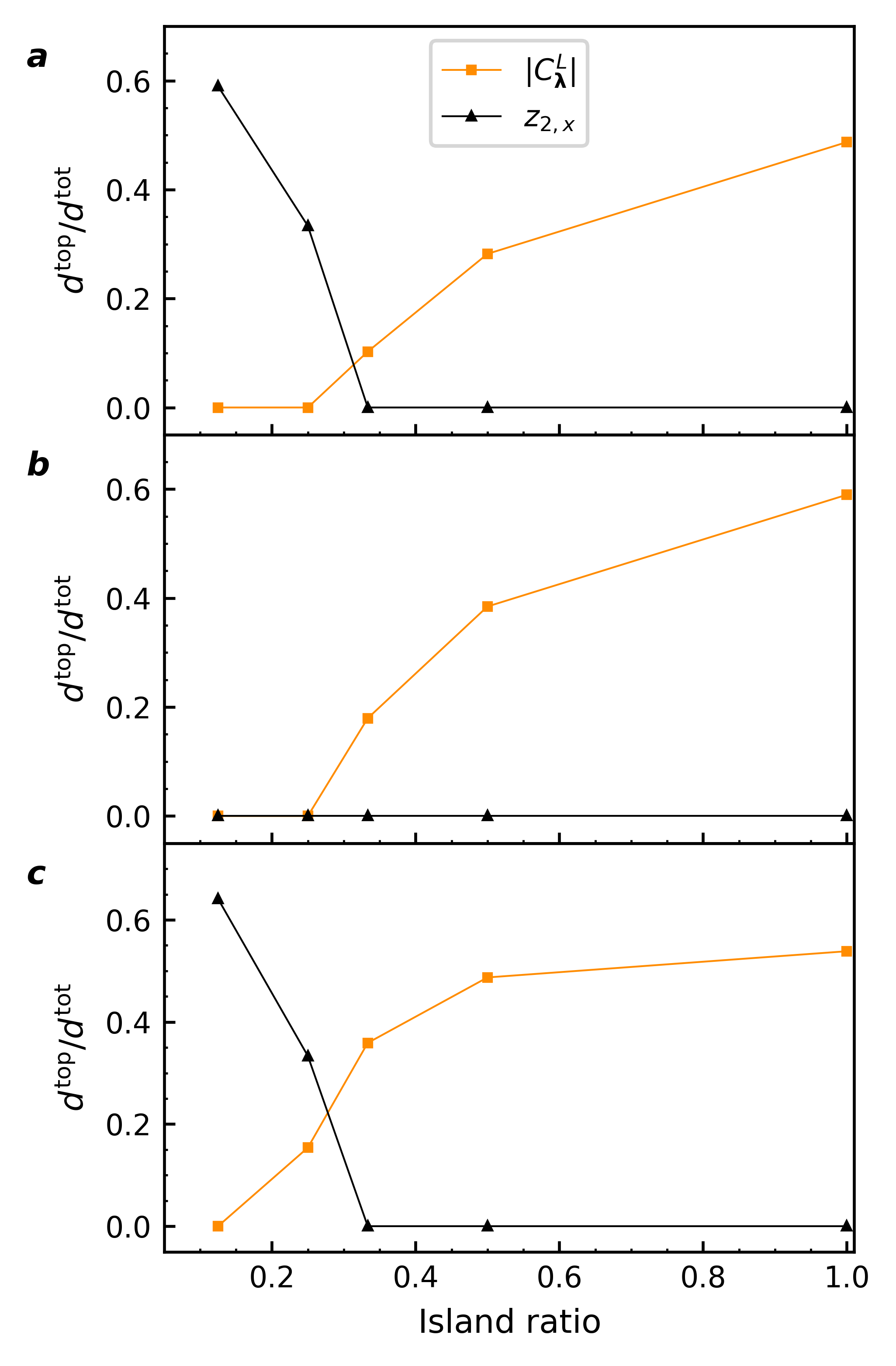}
    \caption{\textbf{Topological dimensional crossover as a function of island width.} Length of the cut along $x$ for $y=N/2$ and $E=0$, as shown in Fig.~\ref{fig:fig2}, for which the local topological invariants $C^{\textrm{L}}_{\boldsymbol{\lambda}}$ (orange squares) and $z_{2,x}$ (black triangles) are nontrivial and localizer gap of appropriate dimension is larger than $0.02t$, $d^{\textrm{top}}$, divided by the number of sites along the cut, $d^{\textrm{tot}}$. \textbf{a} Lattice uses the parameters from Fig.~\ref{fig:fig2}. \textbf{b} Lattice uses the parameters $N=20$, $\mu=0$, $\alpha=0.8t/l$, $\Delta=1.2t$, and $JS=2t$. The set of plots showing the topological transition can be found in Supplementary Note 5. \textbf{c} Lattice uses the parameters from Fig.~\ref{fig:fig3}. The spectral localizer calculations use $\kappa = 0.05 (t/l)$. 
    }
    \label{fig:fig4}
\end{figure}

To understand the various phases revealed in Fig.~\ref{fig:phase-diagram}, we study the topological invariants as the Shiba geometry transitions from two-dimensional island to a one-dimensional chain by varying the distribution of magnetic adatoms placed atop the underlying superconductor.

To begin, we consider the case where both the Shiba island and chain are topological with $|C_{\boldsymbol{\lambda}}^\textrm{L}| = 1$ and $z_{2,x} = -1$, respectively, see Fig.~\ref{fig:fig2}. Specifically, our goal is to connect the change in the distribution of the magnetic adatoms [Fig.~\ref{fig:fig2}a-e] with the spectra of the associated spectral localizers [Fig.~\ref{fig:fig2}f-j], as well as the local markers and associated localizer gaps [Fig.~\ref{fig:fig2}k-t], as the adatom distribution is interpolated between a Shiba circular island and a Shiba chain. For the Shiba island, Fig.~\ref{fig:fig2}f, $\Sigma(L^{(2D)}_{\boldsymbol{\lambda}}(\boldsymbol{X},H))$ crosses zero at the edge of the magnetic adatom island, transitioning from trivial $C_{\boldsymbol{\lambda}}^\textrm{L}=0$ to topological with $|C_{\boldsymbol{\lambda}}^\textrm{L}|=1$ [Fig.~\ref{fig:fig2}k]. The finite localizer gap is maximal at the center of the island. However, for this same geometry, the spectrum of the 1D spectral localizer, $\Sigma(L^{(1D)}_{\boldsymbol{\lambda}}(\boldsymbol{X},H))$, displays a much smaller localizer gap within the Shiba island [Fig.~\ref{fig:fig2}p]. While there are sites where the topological local marker is topological, $z_{2,x} = -1$, the vanishing localizer gap means that there is little protection against variations in the Hamiltonian $\delta H$ or position operators $\delta X$. Therefore, the 1D topology is not robust, and the system is characterized by a two-dimensional topology via the local Chern number, $C_{\boldsymbol{\lambda}}^\textrm{L}$.

As the number of magnetic adatoms is reduced along the y-direction, flattening the island, as shown in Fig.~\ref{fig:fig2}b, similar trends in the local markers and the associated localizer gaps are obtained. However, the number of sites with a nontrivial Chern number is reduced compared with the circular island's case. Additionally, the 2D localizer gap is now smaller. The trend continues for a flatter island, as shown in Fig.~\ref{fig:fig2}c. However, for the parameters considered, the three-atom-wide island exhibits a change in its behavior: the Chern number is now nonzero only at locations where the 2D localizer gap vanishes, Fig.~\ref{fig:fig2}n, while there are sites with a nontrivial $z_{2,x}$ invariant protected by a 1D localizer gap,  Fig.~\ref{fig:fig2}s. As such, this marks the transition from 2D to 1D topology in the system due to the change in the system's geometry. For the Shiba chain, panel (e), the $z_{2,x}$ invariant changes from trivial to nontrivial at the edge of the chain. As the Shiba island reduces to a chain, $\Sigma(L^{(2D)}_{\boldsymbol{\lambda}=(x,y=N/2,E=0)}(\boldsymbol{X},H))$ approaches $\Sigma(L^{(1D)}_{\boldsymbol{\lambda}=(x,E=0)}(\boldsymbol{X},H))$. The local density of states for all the geometries considered is shown in the Supplementary Note 4.

To demonstrate that the topological transition shown in Fig.~\ref{fig:fig2} is robust regardless for different boundaries of adatoms (so long as additional symmetries are not introduced), we consider the dimensional crossover using a geometry without symmetries, as shown in Fig.~\ref{fig:random}. To define the asymmetric Shiba islands, we start from the equation of a circle and define a ``local'' radius by introducing a random variable that changes the distance from the center of the island to the edge by up to $10 \%$. The same procedure is applied to the flattened geometries. For the Shiba lattices that are 2D-like structures, Fig.~\ref{fig:random}a,b, the topology is characterized by the local Chern number, $|C^L_{\boldsymbol{\lambda}}|$ and protected by the large 2D localizer gap. Similar to Fig.~\ref{fig:fig2}, for these 2D geometries, the 1D localizer gap is too small to protect a meaningful $z_{2,x}$. Conversely, as the randomized-boundary system approaches the Shiba chain limit, Fig.~\ref{fig:random}c-e, it becomes trivial from the 2D topological invariant perspective, while displaying regions with a nontrivial $z_{2,x}$ protected by a 1D localizer gap.

We now consider a set of parameters that leads to a topological island with $|C_{\boldsymbol{\lambda}}^\textrm{L}|=2$ transitioning into a topological chain $z_{2,x} = -1$, see Fig.~\ref{fig:fig3}. For this case, there are two main differences with respect to the previous example we discussed. First, there is a topological transition from $|C_{\boldsymbol{\lambda}}^\textrm{L}|=2$ to $|C_{\boldsymbol{\lambda}}^\textrm{L}|=1$ when the Shiba island shrinks from seven to five atoms wide in the $y$ direction. Second, for the three-atom-wide island, there is a coexistence of non-trivial 2D and 1D topology with $|C_{\boldsymbol{\lambda}}^\textrm{L}|=1$ and $z_{2,x} = -1$, respectively, though both topological phases are only weakly protected due to the relatively small localizer gaps in both cases. 

In Supplementary Note 5, we show plots for a topological island with $|C_{\boldsymbol{\lambda}}^\textrm{L}|=2$ transitioning into the trivial chain, $z_{2,x} = 1$. In this case, the number of sites with nontrivial topology decreases together with the localizer gap protection, until the system becomes trivial in both 1D and 2D. Finally, in Supplementary Note~6, we consider the opposite case of a trivial Shiba island with $|C_{\boldsymbol{\lambda}}^\textrm{L}|=0$ transitioning into a topological chain with $z_{2,x} = -1$. In this case, the topological phase transition due to dimensional crossover occurs for the three-atoms-wide island. 

To conclude the discussion of topological phase transitions due to dimensional crossover in the clean limit, we consider a different approach to understanding these transitions based on the number of sites in the system exhibiting a non-trivial local marker with at least a moderate localizer gap, see Fig.~\ref{fig:fig4}. In particular, we consider the number of sites with a topological local marker, $d^{\textrm{top}}$, divided by the total number of sites on the $x$-axis, $d^{\textrm{tot}}$, both for 1D and 2D, as a function of the island's width-to-length ratio. Here, we only consider sites topological if the localizer gap is larger than $0.02t$. Thus, the transition from 2D topology to 1D topology, as characterized by the spectral localizer framework, is approximately given by the crossing of the traces.

\begin{figure}
    \centering
    \includegraphics[width=1\linewidth]{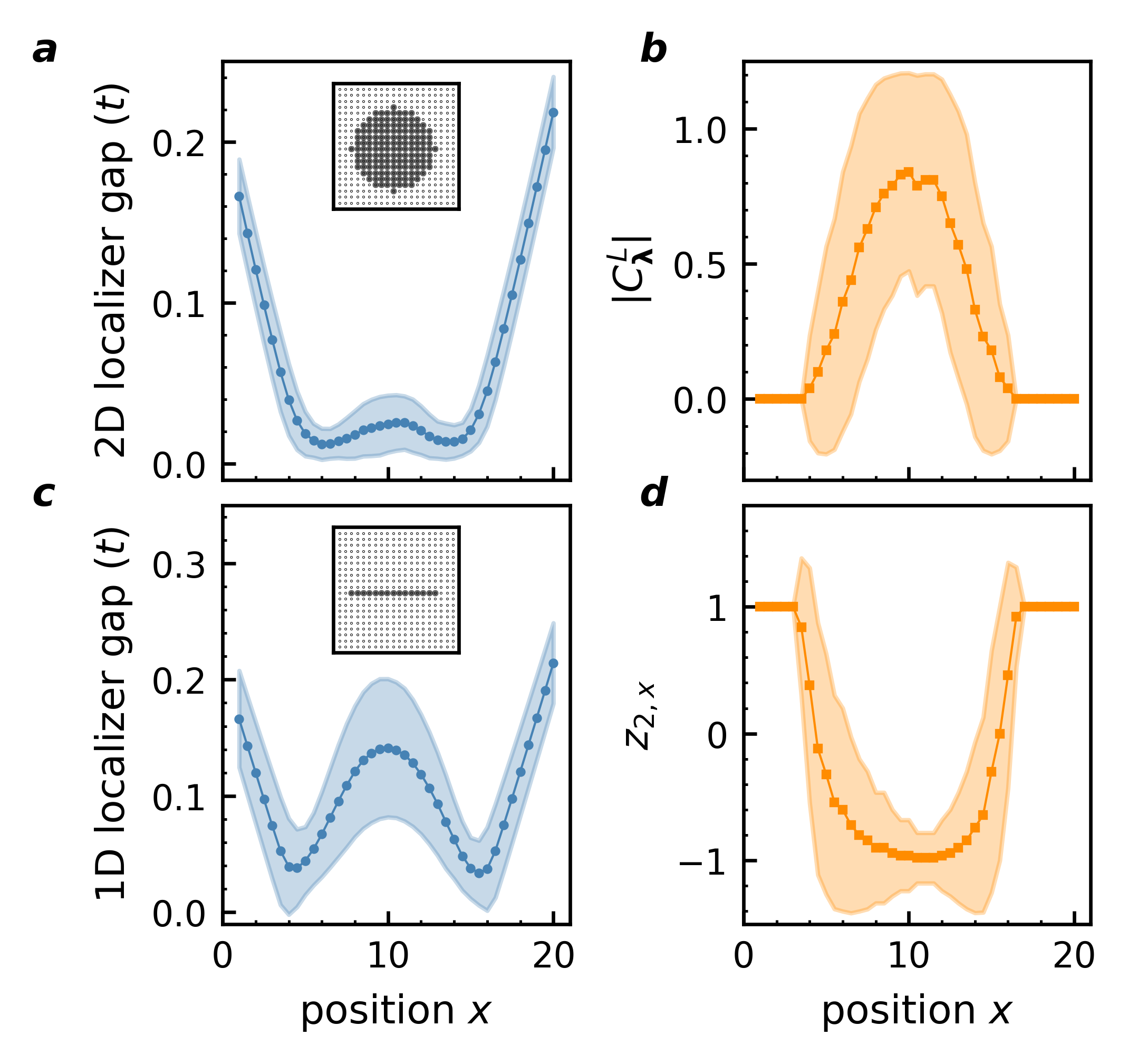}
    \caption{\textbf{Effects of disorder on the Shiba island and chain.} \textbf{a} Ensemble-averaged 2D localizer gap for a disordered circular Shiba island, as shown in the inset with filled and empty circles indicating magnetic adatom locations and absences, respectively, along $x$ for $y=N/2$ and $E=0$, and with disorder strength $W/t=1.2$. The blue dots show the average over 100 disorder realizations, while the light blue shaded area corresponds to $\pm 1$ standard deviation of the average. \textbf{b} Ensemble-averaged local maker $|C_{\boldsymbol{\lambda}}^\textrm{L}|$ (orange squares) and the region falling within $\pm 1$ standard deviation (light orange). \textbf{c} Ensemble-averaged 1D localizer gap (blue dots) and region falling within $\pm 1$ standard deviation (light blue) for a disordered Shiba chain, as shown in the inset, along $x$, for $E=0$ and $W/t=1.2$, corresponding to the chain's position. \textbf{d} Ensemble-averaged local marker $z_{2,x}$ (orange squares) and the region falling within $\pm 1$ standard deviation (light orange). The spectral localizer calculations use $\kappa = 0.05 (t/l)$.
    }
    \label{fig:disorder1}
\end{figure}

\subsection{Effects of disorder}

To test the robustness of our results, we consider the effects of adding disorder in the local chemical potential. To do so, we include an on-site disorder term in the Hamiltonian,
\begin{equation}
    H_{\textrm{dis}} = - \sum_{\mathbf{r},\sigma} \delta \mu(\mathbf{r}) c^\dagger_{\mathbf{r}\sigma} c_{\mathbf{r}\sigma},
\end{equation}
where $\delta \mu(\mathbf{r})$ is taken to be an uncorrelated, uniformly distributed random variable, $\delta\mu(\mathbf{r}) \in [-W, W]$. Shiba lattices could also present disorder in the adatom coupling strength, spin orientation, and possibly missing magnetic atoms in a given geometry \cite{PhysRevB.100.235102}. In this study, we do not calculate the superconducting order parameter self-consistently in the presence of disorder \cite{PhysRevB.95.184511}. Here, we consider the disorder strength $W/t=1.2$ and compute the topological markers and localizer gaps for an island and a chain geometry, averaging over an ensemble of 100 disorder realizations see Fig.~\ref{fig:disorder1}. For this disorder strength, the 2D localizer gap is closed within the Shiba island along the path considered. Thus, whatever local 2D marker the system may possess, the associated phase is not robust. However, the 1D localizer gap remains open and system remains 1D-topological, see Fig.~\ref{fig:disorder1}d.  

\begin{figure}
    \centering
    \includegraphics[width=1\linewidth]{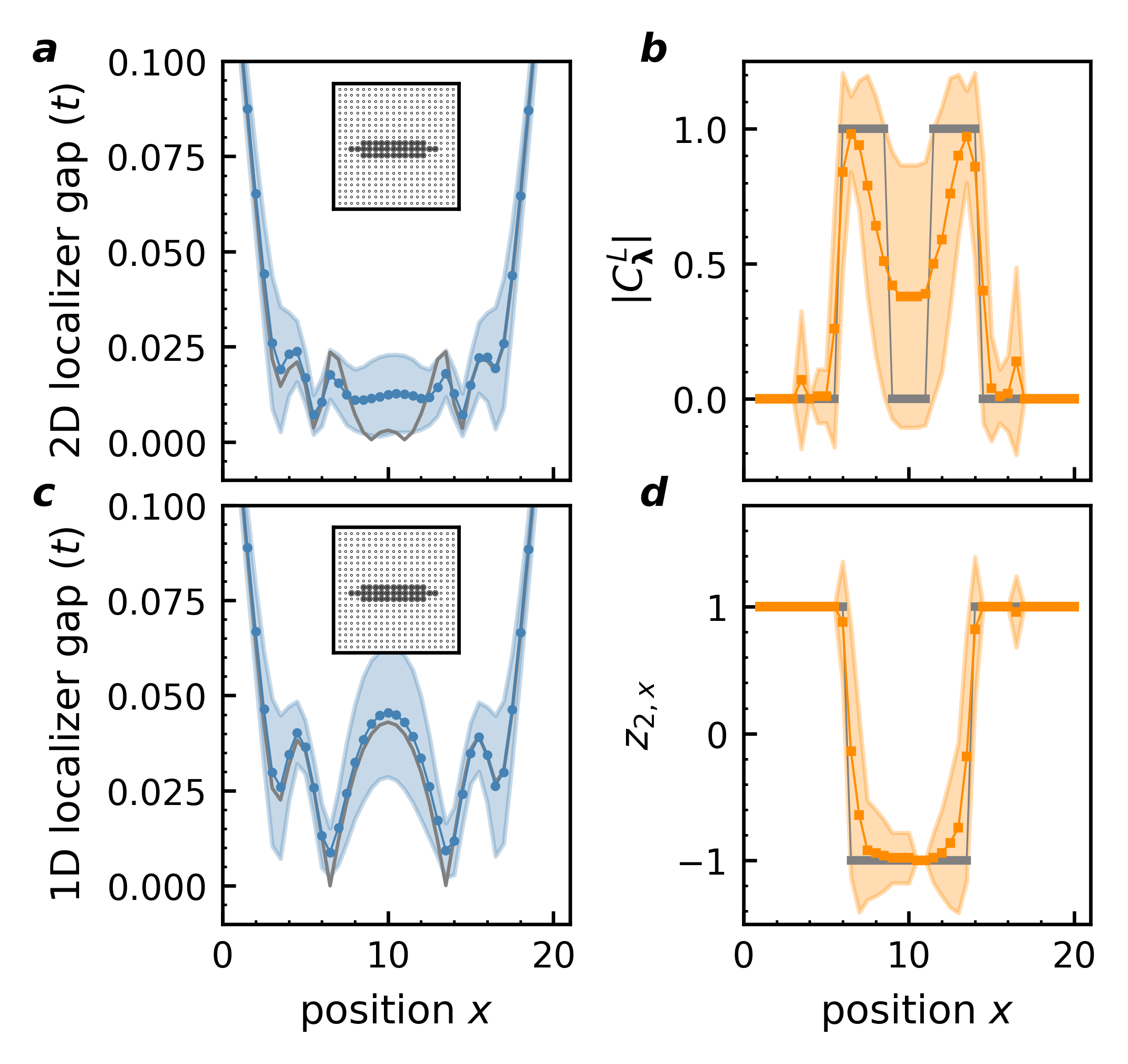}
    \caption{\textbf{Effects of disorder at the topological dimensional crossover point.} \textbf{a} Ensemble-averaged 2D localizer gap for a disordered three-atom-wide Shiba island, as shown in the inset with filled and empty circles indicating magnetic adatom locations and absences, respectively, along $x$ for $y=N/2$ and $E=0$, and with disorder strength $W/t=0.3$. The blue dots show the average over 100 disorder realizations, while the light blue shaded area corresponds to $\pm 1$ standard deviation of the average. \textbf{b} Ensemble-averaged local maker $|C_{\boldsymbol{\lambda}}^\textrm{L}|$ (orange squares) and the region falling within $\pm 1$ standard deviation (light orange). \textbf{c} Ensemble-averaged 1D localizer gap (blue dots) and region falling within $\pm 1$ standard deviation (light blue) for a disordered Shiba chain along $x$, for $E=0$ and $W/t=0.3$, corresponding to the chain's position. \textbf{d} Ensemble-averaged local marker $z_{2,x}$ (orange squares) and the region falling within $\pm 1$ standard deviation (light orange). In \textbf{a}-\textbf{d}, the gray lines correspond to the clean limit. The spectral localizer calculations use $\kappa = 0.05 (t/l)$.
    }
    \label{fig:disorder2}
\end{figure}

Now, we consider the effect of disorder for a narrow Shiba lattice where 1D and 2D topologies coexist in the clean case, as considered in Fig.~\ref{fig:fig3}n,s. In particular, we study a disordered three-atom-wide Shiba island with disorder strength $W/t = 0.3$ and average over an ensemble of 100 disorder realizations, see Fig.~\ref{fig:disorder2}. When considering the system's 2D topology, the small gap protecting its topological phase in the clean limit is now significantly reduced in the presence of disorder. However, again, the 1D localizer gap remains open across the region with nontrivial $z_{2,x} = -1$. Together, these results indicate the chemical potential disorder favors the 1D topology. 

\section{Discussion}

We have established a real-space approach based on the spectral localizer framework to characterize topological transitions due to spatial dimension changes. In particular, we have leveraged local topological markers for systems in class D, with particle-hole symmetry and broken time-reversal and chiral symmetry, to study Shiba lattices in experimentally relevant examples. We considered a 2D circular Shiba island and deformed it into a 1D chain by reducing its size along one of its dimensions. By computing the local markers across the dimension change, we found that the transition from 2D topology to 1D topology occurs via a reduction in the number of sites with Chern number $|C_{\boldsymbol{\lambda}}^\textrm{L}|>0$ together with a decrease in the 2D localizer gap protecting the Chern number. At a critical thickness, we found the emergence of nontrivial 1D topology, marked by  $z_{2,x}=-1$, protected by a nonzero 1D localizer gap whose size increases for the 1D chain geometry. We tested the robustness of our results against disorder in the chemical potential of the magnetic adatoms, and numerically observed that the 1D localizer gap, and thus the 1D topology, is more robust than its 2D counterpart.  

There are other methods to produce local markers of $K$-theory, including some that are valid for interacting systems \cite{markov2021local,hannukainen2024interacting}.  These are, however limited to the $\mathbb{Z}$-valued invariants so cannot be used to track the same Altland-Zirnbauer symmetry class across adjacent dimensions where at least one $\mathbb{Z}_2$-valued invariant will be needed.

The approach we have introduced can be readily generalized to other symmetry classes. For example, we also considered the one-to-zero-dimensional crossover in a BDI class model, showing the generality of our approach [see Supplementary Note 1]. The spectral localizer, Eq.~\eqref{eq:Ldef}, can also be defined for a 3D topological insulator and a 2D quantum spin Hall system \cite{PhysRevB.81.115407}. Additionally, one could consider the simultaneous removal of two or more dimensions, for example, to study the topological transition between 3D and 1D class AIII systems. More generally, the framework could help guide the design of topological materials for device applications. One of the strengths of the spectral localizer framework is its numerical efficiency, which enables it to be applied directly to experimental systems without developing a low-energy approximation \cite{PhysRevLett.134.126601}. As such, our results may be useful for the design and analysis of topological superconducting nanowires \cite{doi:10.1126/science.aav3392}.

\section{Methods}

\subsection{Local density of states}

We compute the local density of states using the Chebyshev-Bogoliubov–de Gennes method introduced in reference \cite{PhysRevLett.105.167006}. For completeness, we provide here a brief description. The Green’s function is defined as $G = \langle vac| \Psi^\dagger \hat G(E) \Psi |vac \rangle$, where $\hat G(E) = \left[E + i \epsilon -  H \right]^{-1}$, $H$ the Hamiltonian in Eq. \eqref{eq:H}, $|vac \rangle$ is the vacuum, and $\Psi = ( c_{1\uparrow}, ..., c_{N^2\uparrow}; c_{1\downarrow}, ..., c_{N^2\downarrow};c^\dagger_{1\uparrow}, ..., c^\dagger_{N^2\uparrow}; c^\dagger_{1\downarrow}, ..., c^\dagger_{N^2\downarrow})$, labeling the positions in the two-dimensional superconductor using a single index. $G^{\sigma \sigma}_{jj}$ in Eq. \eqref{eq:ldos} corresponds to the matrix entry with spin $\sigma$ at site $j$. As shown in reference \cite{PhysRevLett.105.167006}, expanding the Green’s function using Chebyshev polynomials leads to 
\begin{align}\nonumber
G_{jj}^{\sigma\sigma}(E)& \approx \frac{-2 i}{\sqrt{1-\tilde{E}^2}} \sum_{m=0}^M a_m^{\sigma\sigma}(j,j) \frac{\sinh[\nu(1-\frac{m}{M})]}{\sinh(\nu)}\\&
\quad \times e^{-i m \arccos(\tilde E)},
\label{eq:gapprox}
\end{align}
with coefficients $a^{\sigma\sigma}_m(i,j)=\langle c_{i\sigma}| T_m(H) |c_{j\sigma}^\dagger \rangle/(1+\delta_{0,m})$, where $ |c_{j\sigma}^\dagger \rangle = c_{j\sigma}^\dagger|vac \rangle$, and $T_m(x)$ are the Chebyshev polynomials of the first kind which can be computed recursively. $\tilde E$ corresponds to a rescaled energy, such that the spectrum lies within the interval $[-1,1]$. Eq. \eqref{eq:gapprox} becomes exact in the limit $M \rightarrow \infty$. $\nu = \epsilon M$ is a real number taken to be the product of the Green's function broadening $\epsilon$ and the summation cutoff $M$. We consider $\epsilon = 5\times10^{-3}$ and a cutoff $M=10^3$, enough for the system sizes we consider.

\subsection{Spectral localizer}

On the tight-binding basis of the Hamiltonian $H$, given by  $\Psi = ( c_{1\uparrow}, ..., c_{N^2\uparrow}; c_{1\downarrow}, ..., c_{N^2\downarrow};c^\dagger_{1\uparrow}, ..., c^\dagger_{N^2\uparrow}; c^\dagger_{1\downarrow}, ..., c^\dagger_{N^2\downarrow})$, we label the positions in the system using a single index, $j=1,...,N^2$. Then, the position operators $X, Y$ used in the spectral localizer are diagonal matrices with the $x$ ($y$) component of the $j$-th lattice site at the entry $X_{jj}$ ($Y_{jj}$). The spin and particle-hole spaces are labeled with an additional index at the same position. The spectral localizer is diagonalized using eigenvalue solvers to obtain its spectrum and gap. The two-dimensional topological marker, $C^L_{\boldsymbol{\lambda}}$, is given by the signature of the spectral localizer $L^{(2D)}_{\boldsymbol{\lambda}}$. When the full spectrum is not needed, we speed up its calculation using the LDLT decomposition of the spectral localizer. To compute the sign of the determinant needed for the 1D topological marker, $z_{2,x}$, we perform a LU factorization of $L^{(1D)}_{\boldsymbol{\lambda}}$.  

\section*{Data availability}
The data that support the findings of this work are available from the corresponding author on reasonable request.

\section*{Code availability}
The codes that support the findings in this study are available from the corresponding author upon reasonable request.


\section*{Author contributions}
All authors conceived the research project. M.R-V.\ performed the numerical calculations, with input from A.C.\ and T.A.L. All authors contributed to discussions and interpretation of the results. M.R-V.\ and A.C.\ wrote the manuscript with the comments from T.A.L.

\begin{acknowledgments}

We thank Annica M Black-Schaffer and Arnob Kumar Ghosh for insightful discussions. A.C.~acknowledges support from the Laboratory Directed Research and
Development program at Sandia National Laboratories.
This work was performed, in part, at the Center for Integrated Nanotechnologies, an Office of Science User Facility operated for the U.S.~Department of Energy (DOE)
Office of Science. Sandia National Laboratories is a multimission laboratory managed and operated by National Technology $\&$ Engineering Solutions of Sandia, LLC, a
wholly owned subsidiary of Honeywell International, Inc.,
for the U.S.~DOE’s National Nuclear Security Administration under contract DE-NA-0003525.
Research by T.L.~was sponsored by the Army Research Office and was accomplished under Grant Number W911NF-25-1-0052. The views and conclusions contained in this document are those of the authors and should not be interpreted as representing the official policies, either expressed or implied, of the Army Research
Office or the U.S.~Government. The U.S.~Government is authorized to reproduce and distribute reprints for Government purposes notwithstanding any copyright notation herein.
The views expressed in the article do not necessarily represent the
views of the U.S.~DOE or the United States Government.
\end{acknowledgments}

\section*{Competing interests}
The authors declare no competing interests.

\section*{References}

\end{document}